\documentclass[article]{elsarticle}

\usepackage{lineno,hyperref}
\modulolinenumbers[5]

\journal{Carbon}

\usepackage{graphicx}
\usepackage{amssymb}
\usepackage{amsfonts}
\usepackage{amsbsy}

\newcommand{\be}{\begin{equation}}
\newcommand{\ee}{\end{equation}}
\newcommand{\ben}{\begin{eqnarray}}
\newcommand{\een}{\end{eqnarray}}

\begin{document}

\begin{frontmatter}

\title{Optical vortex driven charge current loop and optomagnetism in fullerenes}

%
\author{J. W\"{a}tzel, Y. Pavlyukh, A. Sch\"affer, and J. Berakdar$^1$}
\address{Institut f\"ur Physik, Martin-Luther Universit\"at Halle-Wittenberg, 06099 Halle (Saale), Germany}
\fntext[myfootnote]{Corresponding author. E-mail: jamal.berakdar@physik.uni-halle.de, Tel:+49 345552850, Fax: +49 3455527114}


\begin{abstract}
Endohedral molecular magnets, e.g. as realized in fullerenes containing DySc$_2$N, are promising candidates for molecular electronics and quantum information processing. For their functionalization  an ultrafast local magnetization control is essential. Using full ab-initio quantum chemistry calculations we predict the emergence of charge current loops in fullerenes with an associated orbital magnetic moment upon irradiation
  with  weak light vortex  pulses that transfer  orbital angular momentum.  The generated current is controllable by the frequency, the vortex topological charge, and the intensity of the light. Numerical and analytical results show that an ultraviolet vortex femtosecond pulse with an intensity $\sim 10^{13}\,$W/cm$^2$ generates non-invasively nA  unidirectional  surface current  with an associated magnetic field of hundreds $\mu$T at the center of the fullerene.
\end{abstract}

\begin{keyword}
\texttt{fullerenes\sep molecular magnets\sep vortex beams\sep quantum information\sep endohedrals} %
\end{keyword}

\end{frontmatter}


\section{Introduction}
\begin{figure}[t!]
\centering
\includegraphics[width=6cm]{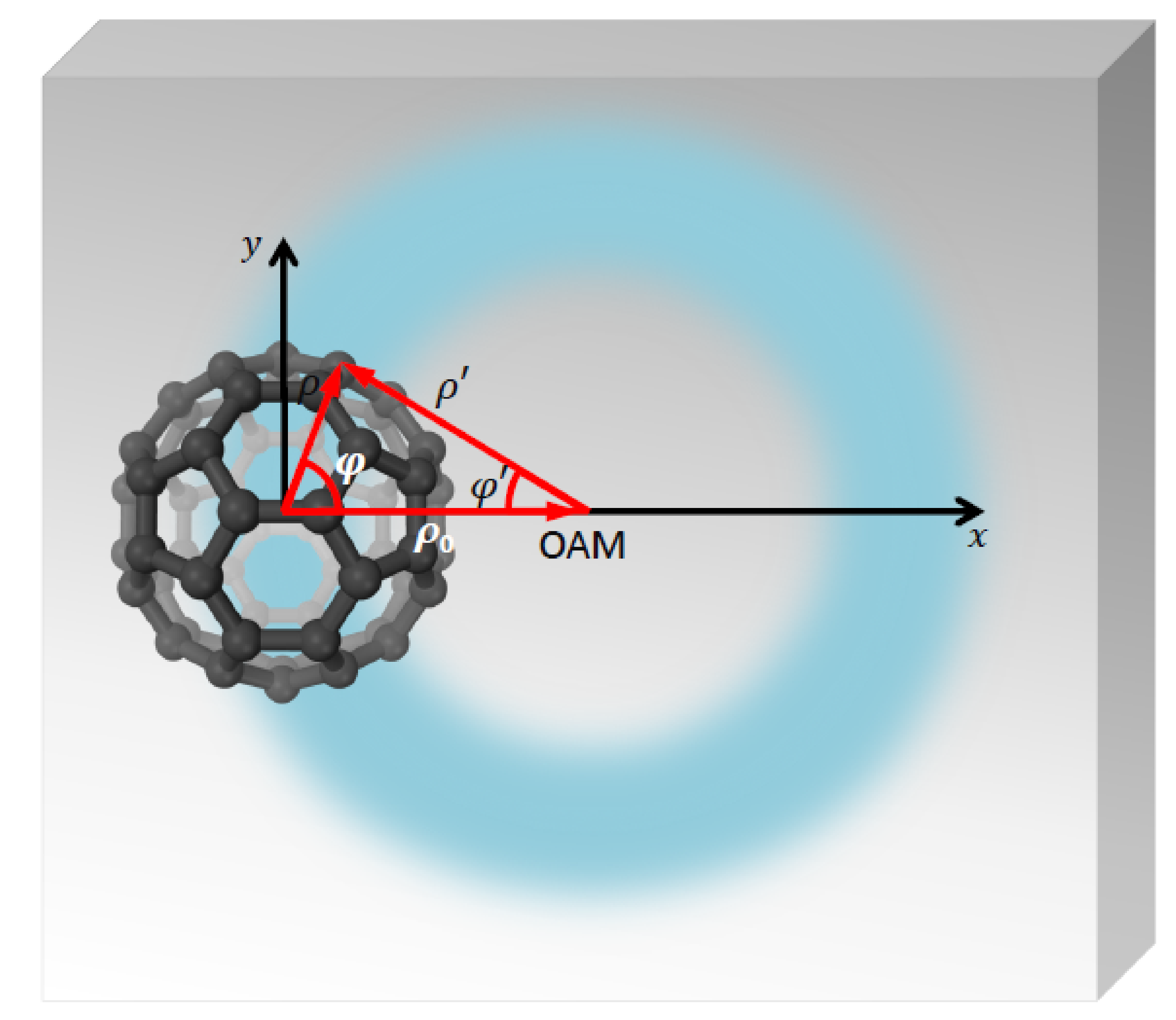}
\caption{A schematic of the coordinate system in the $xy$ plane, perpendicular to the light propagation direction.
 The vector $\boldsymbol{\rho}$  ($\boldsymbol{\rho}_0)$ marks the position of the electron (optical axis) relative to the molecule center of mass. The vector $\boldsymbol{\rho'}$ denotes the position of the electron with respect to the optical axis.  The shaded area is the light high intensity region (not to scale).}
\label{fig:scheme}
\end{figure}

Light sources with a temporal resolution down to femtoseconds are readily available.
In contrast, achieving a spatial resolution relevant for nanostructures, i.e. below the optical wave length, is  a challenge. One way is to utilize
the light-matter interaction to ignite and tune carriers dynamics. For instance, a triggered ultrafast charge current nano loop generates an equally localized magnetic field that   steers selectively the properties of  magnetic nanoparticles.
As demonstrated below, such a scenario  is realized  by  an optical vortex beam triggering a valence charge current loop on a fullerene sphere. These fullerenes, as for instance DySc$_2$N@C$_{80}$\cite{jacs12}, may encapsulate endohedral molecular magnets that are shielded and have thus a long relaxation time which  makes them good candidates for sensing and (quantum) information applications.  Gd$_3$N@C$_{80}$ \cite{prl13} and N@C$_{60}$ \cite{nmat09} are further experimentally feasible magnetic endofullerenes with interesting applications\cite{prl13,nmat09}. The present study relies on the idea of transferring light orbital angular momentum to bound  degenerate states  in fullerenes leading to the emergence of a magnetic orbital moment.
We evaluate then the magnetic field  inside the fullerene cage due to the respective surface charge currents.
To this end two  established facts are important:\\
\emph{i)} Recent experiments \cite{petek} and first principle calculations \cite{PavlyukhAngular2009,pavlyukhkohn2010,huang2010superatom,pavlyukhCommunication2011,
feng2013energy,voora2013existence,klaiman2014best,zakrzewski2014electron}
unveiled the existence of so-called super atomic molecular orbitals (SAMOs) in C$_{60}$ which are diffuse, \emph{unoccupied}  orbitals \emph{bound} by the (molecular cage) central potential and have thus well-defined  orbital angular momenta ($s, p$, and $d$); they  extend way beyond the site-defined bound $\pi$ orbitals and hence they are less affected by doping or modifications of the geometry. Indeed, \emph{ab initio} calculations suggest the existence of SAMOs for larger  or doped fullerenes. Even, for various  bowl-shaped fragments of $\pi$-conjugated molecules SAMOs are present \cite{pccp15}. Here we exploit these properties of SAMOs for photo-generating surface  charge current loops, and hence internal, nanoscale magnetic field pulses which act on encapsulated magnetic moments, e.g.  N@C$_{60}$ \cite{nmat09}.
 As our ab-initio calculations explicitly demonstrate
 N@C$_{60}$ \cite{nmat09} possess indeed SAMOs that we may employ for magnetic pulse generation  (cf. supplementary materials).\\
\emph{ii)} Special type of structured light pulses will be used, namely optical vortices with a topological charge that sets the amount of orbital angular momentum (OAM) transferrable when interacting with matter. Such OAM beams have found an impressively  wide range of applications \cite{AllenOrbtial1992,BabikerLight1994,molina2007twisted,allen1999orbital,mair2001entanglement,
padgett2011tweezers,furhapter2005spiral,torres2011twisted,andrews2011structured}. Of particular relevance  here are the recently generated fs OAM pulses \cite{OAMpulses} and OAM sources in the VUV and EUV regimes \cite{OAMx1,OAMx2,OAMx3}. Due to  phase singularity, the intensity of the OAM beam vanishes on the optical axis and rises to form a ring shape with a radius determined by the beam's waist (cf. Fig.\ref{fig:scheme}). The advantage of pumping current-carrying SAMOs  with linear polarized OAM beam is that one is not restricted to the conventional optical propensity rules.  So the  current (and therefore the magnetic moment) can be enhanced by increasing the topological charge for fixed moderate pulse intensity,
 as demonstrated below. Furthermore, for larger systems, e.g. ring structures \cite{RingOAM1,RingOAM2}, one may use the OAM spatial structure to couple locally to the system (i.e., only to the sphere region but not to the endohedral objects). As schematically shown in Fig.\ref{fig:scheme}, in the case of C$_{60}$  the extension of the diffraction-limited beam is much larger than that of the cage.
  Hence,  it is relevant to inspect the  charge-current dependence  on the distance $\rho_0$ between    C$_{60}$ center and the optical axis. Perhaps unexpected, we find a smooth dependence of the magnetic moment on $\rho_0$. The increase of the field intensity with $\rho_0$ counter-acts the amount of OAM  (with respect to the center of the cage) acquired by SAMOs, which means, in a dilute gas-phase collection of C$_{60}$ distributed over the OAM beam spot roughly the same magnetic field is generated in the cage. As demonstrated here, this magnetic field can be tuned in sign and amplitude by varying the properties of the OAM beam, such as the topological charge, the intensity, the waist, or the frequency.

\section{Theoretical model}
We consider a linearly polarized, monochromatic Laguerre-Gaussian OAM pulse with a topological charge $m_{\rm OAM}$ and frequency $\omega$.
The vector potential in cylindrical coordinates with the  $z$-axis chosen along the light propagation (with a wave vector $q_z$) is given by \cite{AllenOrbtial1992}:
\begin{equation}
\boldsymbol{A}(\boldsymbol{r},t)=\boldsymbol{\hat{\epsilon}}A_0 f_{m_{\rm OAM}}^p(\boldsymbol{r}) e^{i(m_{\rm OAM}\varphi'-
\omega t)}\Omega(t)e^{iq_z z} + {\rm c.c.}\, .
\end{equation}
%
$\boldsymbol{\hat{\epsilon}}$ is the polarization vector and  $\Omega(t)=e^{-\delta t^2}$ is the pulse temporal envelope. $\varphi'$ is the polar angle relative to the optical axis of the laser field. Of interest here is the dynamics transversal to $q_z$. For the photon energies $\hbar\omega$ used here as well as for the size of the molecule ($\sim7 {A}^o$) we deduce $q_z z\ll1$, i.e. the dipole approximation is acceptable along the $z$-axis. The radial structure is described by the function
\begin{equation}
f_{m_{\rm OAM}}^p(\rho')=C_{m_{\rm OAM},p}\,e^{-\frac{\rho'^2}{w_0^2}}\left( \frac{\sqrt{2}\rho'}{w_0}\right)^{|m_{\rm OAM}|}L_p^{|m_{\rm OAM}|}\left(\frac{2\rho'^2}
{w_0^2}\right),
\end{equation}
where $\rho'$ is the distance in the $xy$ plane to the optical axis and can be translated to $\boldsymbol{r}$.   $p$ indexes    the radial nodes. The functions $L_p^{|m_{\rm OAM}|}(x)$ are the generalized Laguerre polynomials. For illustration we assume $p=0$, i.e. $L_p^{|m_{\rm OAM}|}(x)=1$. The case of $p\neq 0$ presents no further complications (and adds no further qualitative information). $w_0$ is the beam waist setting the extent of the optical vortex. Highest intensities are around $\rho_{\rm max}=\sqrt{\frac{m_{\rm OAM}}{2}}w_0$. We will employ a diffraction-limited beam implying that $2w_0\approx\lambda$, i.e., in our simulation we used a waist $w_0=50$\,nm to cover up the photon energies $\hbar\omega$ in a range from 5~eV to 18~eV. For a reasonable comparison of the generated magnetic moments as a function of different topological charges we use a normalized amplitude of the vector potentials such that $A^{m_{\rm OAM}}_{\rm max}=A_0\left|m_{\rm OAM}\right|^\frac{|m_{\rm OAM}|}{2}e^\frac{|m_{\rm OAM}|}{2}.$ Thus, the vector potential  normalization is set by $C_{m_{\rm OAM},p=0}=A_0/A^{m_{\rm OAM}}_{\rm max}$.
\\
\begin{figure*}[t!]
\centering
\includegraphics[width=12cm]{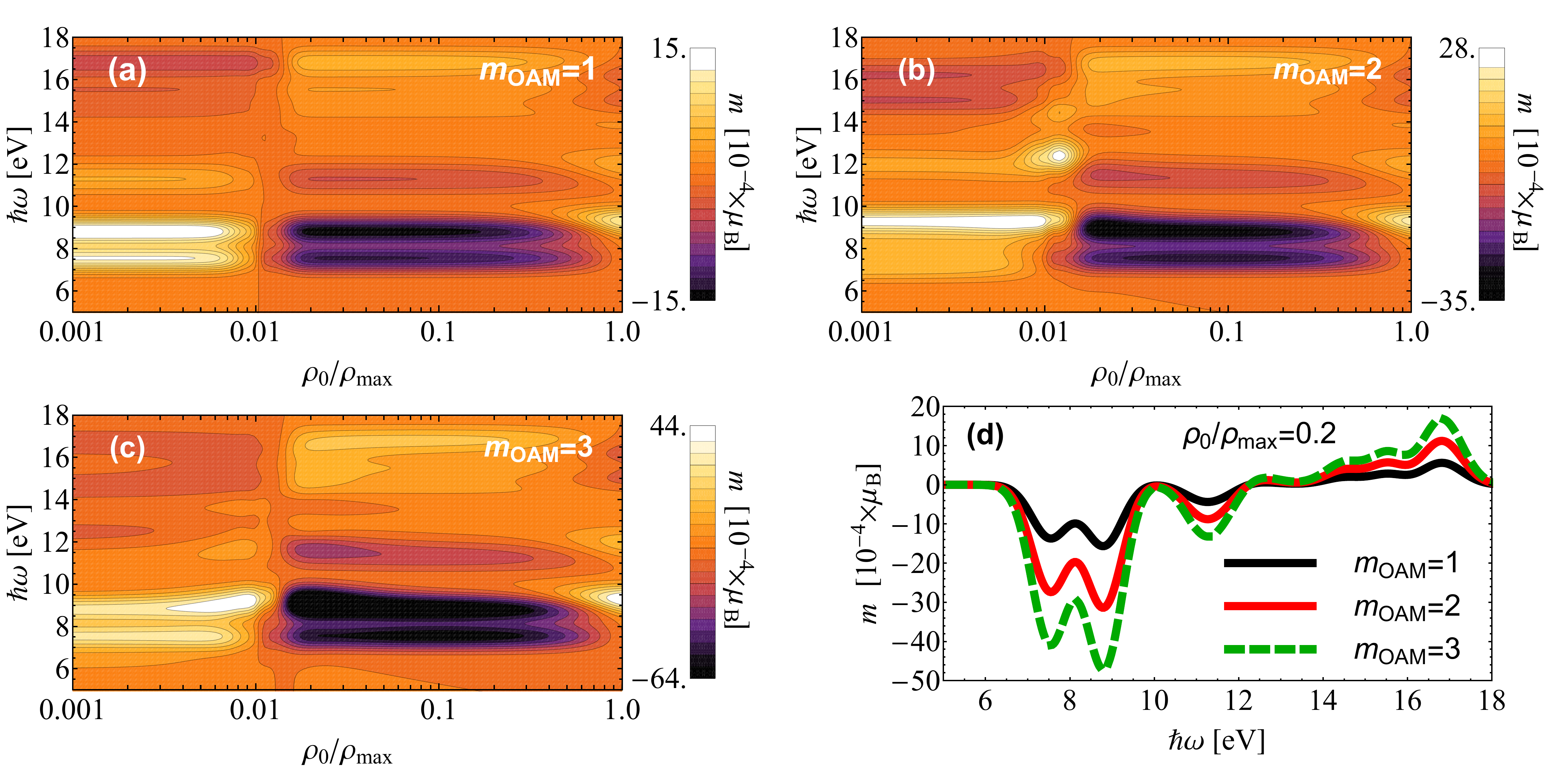}
\caption{(a) Dependence of magnetic moments in $z$-direction   on the photon energy $\hbar\omega$ and the distance between the optical axis and the center of the molecule (given by the ratio $\rho_0/\rho_{\rm max}$) for $m_{\rm OAM}=1$. (b) Same as (a) for $m_{\rm OAM}=2$. (c) Same as (a) for $m_{\rm OAM}=3$. (d) The magnetic moment  variation  with   photon energy for $\rho_0/\rho_{\rm max}=0.2$.}
\label{fig:cur}
\end{figure*}
{As we are interested in non-invasive, direct one-photon nearly resonant generation of
 SAMOs loop current, the choice of the pulse  parameters is a key as issue:
 The beam should have
  a moderate intensity $I$ and  UV-frequencies $\omega$. Its duration should be short but
    allows for few optical cycles. Increasing the generated current is achieved not by increasing the
    pulse intensity but by increasing $m_{\rm OAM}$. The energetic positions of SAMOs are very helpful:
    As $I/\omega$ is the small parameter in time-dependent perturbation theory, for  $\omega$ being in the UV regime, $I$ can
    be chosen relatively high while
      strong field or tunneling effects remain  marginal and only single photon processes contribute.  Furthermore, since $\omega$ is far
      off vibronic as well as plasmonic modes, no heating  nor internal field effects are operational during excitations.
      Thus, by tuning $\omega$ in resonance with SAMOs
      one generates  a non-destructive, relatively long-lived (after pulse)  current loops.
      For the theoretical treatment, the fullerene electronic structure  is the key ingredient.}
 Fullerenes  have $I_h$ point-symmetry. The  geometry is checked against the known coordinates of the inequivalent atoms \cite{PavlyukhAngular2009,pavlyukhCommunication2011,pavlyukhkohn2010}. The calculations of the ground state wave functions and energies were performed using the HF method which is implemented in the GAUSSIAN 03 quantum chemistry package \cite{frisch2004gaussian}. The single particle states of the molecule can be grouped in radial bands characterized by the principal quantum number $n$ and have a well-defined angular quantum number $\ell$. In addition to it they are further specified by $p_{\ell}$ which discriminates between the different representations of the $I_h$ symmetry group of the same angular momentum $\ell$ and $\lambda$ which stands for the element within the multi-dimensional representation $p_{\ell}$. The wave function corresponding to the electron state $i$ is represented by $\Psi_i(\boldsymbol{r})=R_{n_i\ell_i}(r)\sum_{m=-\ell_i}^{\ell_i}C_{\ell_i,m_i}^{p_{\ell_i},\lambda} Y_{\ell_i,m_i}(\Omega_{\boldsymbol{r}})$. The coefficients $C_{\ell_i,m_i}^{p_{\ell_i},\lambda}$ of the symmetry adapted functions belonging to the $p_{\ell_i}$ representation within the $I_h$-symmetry point group are well-known and tabulated \cite{shirai1992basis}. The energy which corresponds to the electron state $i$ is characterized by $\varepsilon_i\equiv\varepsilon_{n_i,\ell_i,p_{\ell_i}}$. The lower band ($n=1$) is occupied by 180 electrons, while we find 60 electrons in the second radial band. For $\ell<3$ the energy dispersion follows nearly perfectly the parabolic dispersion $E_i=E_{n_i}+\ell(\ell+1)/2R^2$, where $R$ is the averaged radius of the molecule as in the $I_h$ symmetry the S,P,D orbitals transform as $a_g$, $t_{1u}$ and $h_g$, respectively. For larger $\ell$ we find a small splitting of the energy levels due to the SO(3)$\rightarrow I_h$ symmetry-breaking. However, this splitting is smaller than 0.5\,eV for the second radial band \cite{PavlyukhAngular2009}. The SAMOs are bound virtual states occupying the third radial band ($n=3$). In contrast to their symmetry and radial structure (which are essential
 for optical transitions), the exact energetic positions of SAMOs turned out to be very delicate to capture theoretically due to correlation effects \cite{voora2013existence,PavlyukhAngular2009,pavlyukhCommunication2011,pavlyukhkohn2010,
klaiman2014best,zakrzewski2014electron}. In the following we consider only transitions between the second radial band with a maximal angular momentum quantum number $L_{\rm max}=5$ and the third radial band filled with SAMOs and characterized by $L_{\rm max}=3$. Consequently the photon energy of the electromagnetic pulse is restricted to a range from 5~eV to 18~eV.\\
Using a gauge in which the scalar potential vanishes, the light matter interaction including  transversal and longitudinal parts reads (unless otherwise stated, we use atomic units and denote the momentum operator as $ \hat{\boldsymbol{p}}$)
$\hat{H}_{\rm int}(t)=-\frac{1}{2}\left[
 \hat{\boldsymbol{p}}\cdot\boldsymbol{A}(\rho',\varphi',t)
+ \boldsymbol{A}(\rho',\varphi',t)\cdot\hat{\boldsymbol{p}}\right].
$ 
For the intensities used here the ponderomotive potential $A^2$ has no influence (on the current anyway). As mentioned,
 single-photon processes are by far dominating, i.e.,  the time dependent wave function $\Psi_i(\boldsymbol{r},t)$ that evolves from an initial state $i$ (characterized by the energy $\varepsilon_i$, the quantum numbers $n_i$ and $\ell_i$ as well as the parameters $p_\ell$ and $\lambda$) under the action of $\hat{H}_{\rm int}(t)$ is obtained by evaluating (to  first order in $A$) the
deviation $\delta\Psi_i(\boldsymbol{r},t)$ from the initial state as an expansion over the unperturbed eigenfunctions $\psi_{j}(\boldsymbol{r})$. I.e., $\delta\Psi(\boldsymbol{r},t)=\sum_jB_j e^{-i\varepsilon_jt}\psi_j(\boldsymbol{r})$.  {Standard  perturbation theory \cite{Fedorov} }yields the occupation probability of the state~$j$ after the pulse from the coefficients $B_j(t)$, where $B_j(t)=iG_{\varepsilon_j,\varepsilon_i}(t)\langle \psi_j|H_{\rm int}|\psi_i\rangle$, $\hat{H}_{\rm int}(t)=H_{\rm int}e^{-i\omega t - \delta t^2},$  and $G_{\varepsilon_j,\varepsilon_i}(t)$ is given by
$G_{\varepsilon_j,\varepsilon_i}(t)=\int_{-\infty}^{t}e^{i(\varepsilon_{j} - \varepsilon_{i}-\omega)\tau-\delta \tau^2}{\rm d}\tau + {\rm c.c.}$.\\
 Let $T_1$ be the time where the pulse is truly off. For times $t_f>T_1$ we find that $G_{\varepsilon_j,\varepsilon_i}(t_f)=G_{\varepsilon_j,\varepsilon_i}(T_1)$ and therefore the coefficients $B_j$ become time-independent.
The  photoinduced current density after the interaction with the pulse ($t_f>T_1$) can be evaluated according to \\ $\boldsymbol{j}(\boldsymbol{r},t_f)=\sum_i^{\rm occ.}\Im\left\{\Psi_i^*(\boldsymbol{r},t_f)\nabla\cdot
\Psi_i(\boldsymbol{r},t_f)\right\}$.  Generically, the current consists of  a
fast oscillating (coherence) current related to the time variation of the oscillating induced dipole and a slowly decaying (population) component,
It is the latter one that generates loop current and hence orbital magnetic moment, as explicitly demonstrated for a quantum ring in Ref.\cite{andrey} where also (ps) relaxation due to phonons were calculated. Here we find the same current characteristics (we do not  calculate the
vibrational relaxation but in \cite{pavlyukhCommunication2011} we showed that  SAMOs  initial decay due electronic processes
 is even slower than HOMO).
Now, suppose  we are interested in current-induced processes, e.g. magnetic switching, with a  time scale ($T_{obs}$), say comparable to the typical current life-time  (effective decay constant $\eta$), i.e. $T_{obs} \eta\simeq 1$.
 The relevant  DC current component is obtained by averaging the current $\boldsymbol{j}(\boldsymbol{r})$ over the fast oscillation
\begin{eqnarray}
\boldsymbol{j}(\boldsymbol{r})&=\sum_{k}^{\rm occ.}\Im \left\{\frac{1}{T_{obs}} \int_{T_1}^{T_1+T_{obs}}\sum_{i,j}^{\rm unocc.} \langle\psi_k|H_{\rm int}|\psi_i\rangle\langle\psi_j|H_{\rm int}|\psi_k\rangle \right.\nonumber\\
&\left.~~~~~~~~~~~~~~~~\times G^*_{\varepsilon_i,\varepsilon_k} G_{\varepsilon_j,\varepsilon_k}e^{i(\varepsilon_i-\varepsilon_j)t}\psi^*_i(\boldsymbol{r}) \nabla\cdot\psi_{j}(\boldsymbol{r})\,{\rm d}t\right\}\nonumber\\
&=\sum_{k}^{\rm occ.}\Im\left\{\sum_{\ell,p_\ell}^{\rm unocc.}\sum_{\lambda,\lambda'} M^*_{n\ell p_{\ell}\lambda,n_k\ell_kp_{\ell_k}\lambda_k} M_{n\ell p_{\ell}\lambda',n_k\ell_kp_{\ell_k}\lambda_k}\right.\nonumber\\
&~~~~~~~~~~~~~~~~\times\left.\left|G_{\varepsilon_{n\ell p_\ell},\varepsilon_k}\right|^2 \psi^*_{n\ell p_{\ell}\lambda}(\boldsymbol{r})\nabla\cdot\psi_{n\ell p_{\ell}\lambda'}(\boldsymbol{r})\right\},
\label{eq:current}
\end{eqnarray}
where $M_{n_j\ell_jp_{\ell_j}\lambda_j,n_i\ell_ip_{\ell_i}\lambda_i}$ is the matrix element $\langle\psi_j|H_{\rm Int}|\psi_i\rangle$. We used that the time integral $\frac{1}{T_{obs}} \int_{T_1}^{T_1+T_{obs}} e^{i(\epsilon_i-\epsilon_j)t} e^{\eta(t-T_1)} dt$ predominantly picks up contributions of the states fulfilling $|\epsilon_i-\epsilon_j|<\eta$. Therefore the sums over substates ($\lambda, \lambda'$) corresponding to the same $n,\ell$ are present in eq.~\ref{eq:current}. Our numerical results show that only the angular components  $j_\varphi$ of the current density vector contributes because for the chosen frequency the ionization channel is suppressed. The magnetic moment induced by the current is calculated as \cite{jackson1962classical}:
$\boldsymbol{m}=\frac{1}{2}\int{\rm d}\boldsymbol{r}\,\left[\boldsymbol{r}\times\boldsymbol{j}(\boldsymbol{r})\right].
$
Since only the angular component $j_\varphi$ contributes to the whole current density the magnetic moment points into the $z$-direction. The current density is located around the shell of the molecule (see below). Hence, the corresponding circulating current in the $xy$-plane induces a magnetic field in the center of the molecule which can be calculated according to the Biot-Savart law \cite{jackson1962classical}:
$B(r=0)=\frac{\mu_0}{4\pi}\int{\rm d}\boldsymbol{r}'\,\boldsymbol{j}(\boldsymbol{r}')\times\frac{\boldsymbol{r}'}{r'^3}.
$ 
\section{Results}
As an illustration we choose a pulse with a FWHM of 10~fs ($\delta=1.6\times10^{-5}$~a.u.) and a beam waist of $w_0=50$~nm. We employ moderate intensities in the range of $3\times10^{13}$~W/cm$^2$.
Figs. \ref{fig:cur}(a)-(c) show the photoinduced magnetic moments in $z$-direction as a function of the fullerene position in the laser spot and the frequency of the linearly polarized OAM pulse (a Gaussian pulse does not lift the degeneracy and hence does not generate any DC current). Probably unexpected, at a fixed frequency, the magnetic moment has the same order of magnitude for fullerenes at different positions in the laser spot.  This is not a generic behavior but due to our system size. The reason is that $m_{\rm OAM}$ is defined with respect to the optical axis (not with respect to the fullerene cage center).
Since $R\ll w_0$ molecules in the beam center ($\rho_0/\rho_{\rm max}=0$) are hardly excited due to the low intensity. On the other side for fullerenes at the beam peak intensity, i.e. $\rho_0/\rho_{\rm max}=1$, the distance between the optical axis and the molecule is very large. Thus, the transferred OAM when translated to OAM with respect to the cage center is quite small. The interplay of those two extremes results in the relatively smooth dependence of the magnetic moment on the fullerenes spatial distribution.
This is also endorsed by the optical selection rules for different $m_{\rm OAM}$ and varying $\rho_0/\rho_{\rm max}$. For a ratio $\rho_0/\rho_{\rm max}=0$ the transitions are governed by the selection rule $\Delta L\leq m_{\rm OAM}+1$ and similar to \cite{PiconPhotoionization2010,PiconTransferring2010,KoksalCharge2012}. As the magnetic quantum number $m$ is not a good quantum number for  fullerene the selection rule is restricted to  $\ell$. For $\rho_0/\rho_{\rm max}\rightarrow1$ the transfer of orbital angular momentum is small since the angle $\varphi'$  is very small [cf. with fig.~(\ref{fig:scheme})]. For $\rho_0\approx\rho_{\rm max}$ the dipole selection rule for the orbital angular momentum quantum number, i.e. $\Delta L=1$ is recovered. In fact the selection rules for $\rho_0\rightarrow0$ and $\rho_0=\rho_{\rm max}$  can be identified in fig.~\ref{fig:cur}(a)-(c) by contrasting with the energy spectrum (which we have done). Fig.~\ref{fig:cur}(d)  shows  the current  dependence on $\hbar\omega$ for $\rho_0/\rho_{\rm max}=0.2$. Obviously the different topological charges $m_{\rm OAM}$ lead to  same transitions. They are characterized by non dipolar contributions, i.e. $\Delta L=0$ and $\Delta L=2$. More details on  various  transitions  are  in the supplemental material \cite{SM}.\\
\begin{figure}[t!]
\centering
\includegraphics[width=11cm]{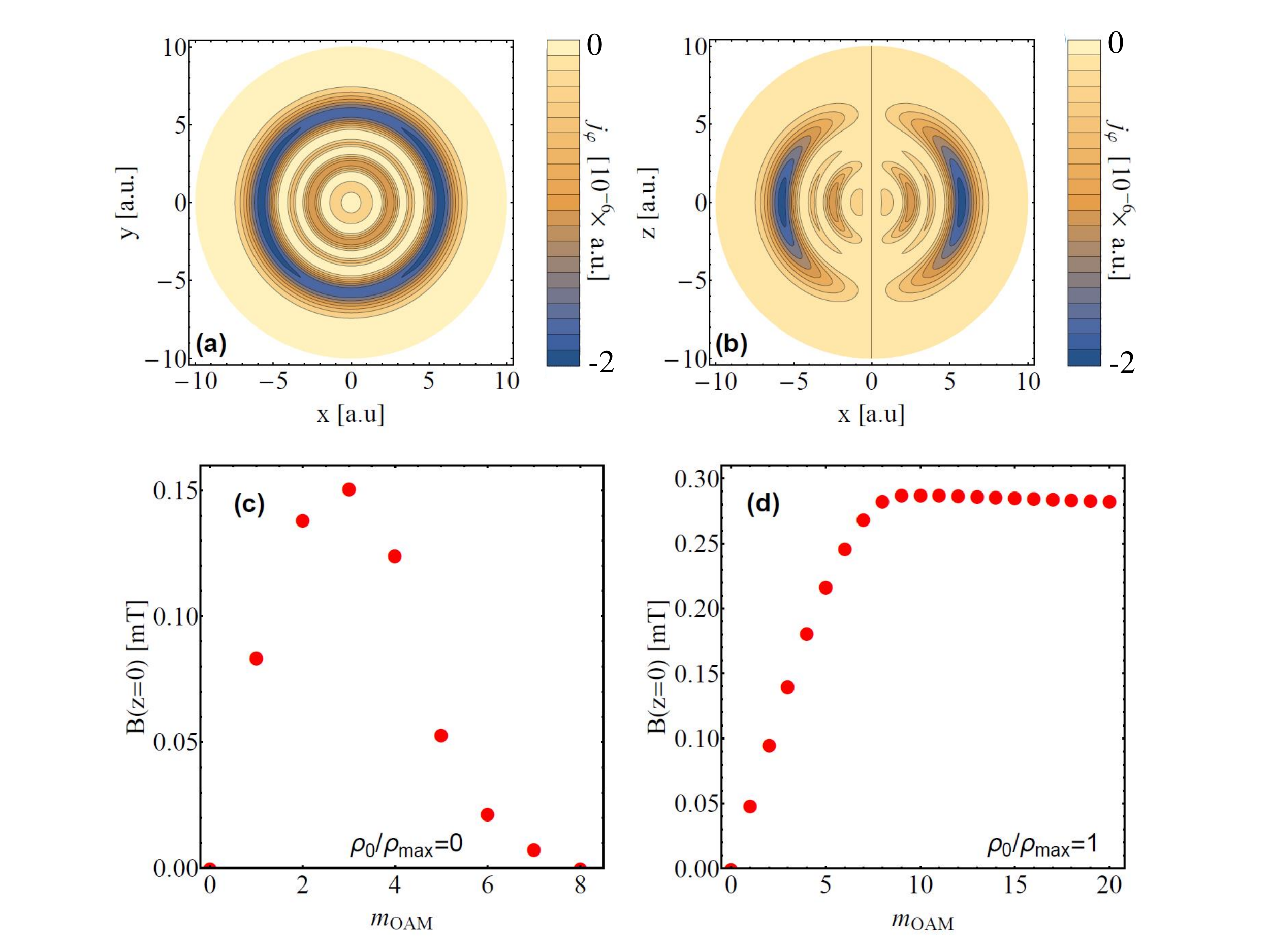}
\caption{(a) Current density in  $xy$ plane for $m_{\rm OAM}=1$, (b) Current density in the $xz$ plane for $m_{\rm OAM}=1$. {The sign of the
 current on the color scale
indicates the direction of the current.} (c) Magnetic field $B$ at the center of the molecule as a function of the topological charge $m_{\rm OAM}$ for $\rho_0/\rho_{\rm max}=0$. (d) Same as (c) but $\rho_0/\rho_{\rm max}=1$.}
\label{fig:Bfield}
\end{figure}
In fig.~\ref{fig:Bfield}(a) and (b) the current densities corresponding to the positive peak of the magnetic moment at 8.8~eV for $\rho_0/\rho_{\rm max}=1$ and $m_{\rm OAM}=1$ are shown. Due to the nodal structure of SAMOs, the current density forms three intensive intercalated ring structures in the $xy$-plane and is nearly independent of the polar angle $\varphi$. The maximal current can be found in the $xy$-plane. The influence of the topological charge is revealed by inspecting the magnetic field magnitude  as  $m_{\rm OAM}$ varies (cf. fig.~\ref{fig:Bfield}(c)  for $\rho_0/\rho_{\rm max}=0$ and  photon energy of 8.8~eV). Not surprisingly we find no generated magnetic field (and no current) for $m_{\rm OAM}=0$, as  for a conventional linearly polarized gaussian beam. The maximal transferable amount of OAM is given by $m_{\rm OAM}=7$, which is because the maximal orbital angular momentum of the third radial band is $L_{\rm max}=3$, i.e. only the transition $\varepsilon_{n_i=2,\ell_i=5}\rightarrow\varepsilon_{n_f=3,\ell_f=3}$ is possible. The maximum of the magnetic field can be found for $m_{\rm OAM}=3$. It reaches strengths in the regime of $\mu$T. {We figured out the individual optical transitions contributing to the current for
a certain position of the fullerene and listed them in the supplementary materials to this work, allowing so to rationalize
all the features observed in (cf. fig.~\ref{fig:Bfield}(c,d) }.
A different picture is obtained in case ($\rho_0/\rho_{\rm max}=1$) in fig.~\ref{fig:Bfield}(d). In general a higher topological charge can be used for current generation. For small winding numbers the generated magnetic field shows a strong dependence, i.e. by changing $m_{\rm OAM}$ from 1 to 2 the field strength is doubled. However, for values above 10 we find a saturation effect, due to the limited number of degenerate virtual states in the third radial band of the C$_{\rm 60}$ molecule, meaning that the amount of the maximally  transferrable orbital momentum is also set by the band structure. Hence, OAM beams for generating magnetic pulses come to their full advantage when $m_{\rm OAM}$ is large and the virtual states are highly degenerate. The fact that SAMOs seem to be ubiquitous \cite{voora2013existence,PavlyukhAngular2009,pavlyukhCommunication2011,pavlyukhkohn2010,
klaiman2014best,zakrzewski2014electron,huang2010superatom,feng2013energy,pccp15} for $\pi$-conjugated molecular and large structures, and very high $m_{\rm OAM}$ is realizable  \cite{AllenOrbtial1992,BabikerLight1994,molina2007twisted,allen1999orbital,mair2001entanglement,
padgett2011tweezers,furhapter2005spiral,torres2011twisted,andrews2011structured}, underlines the promising potential for OAM-photo-assisted current generation in matter.
\section{Conclusions} We studied theoretically the possibility of generating internal charge currents in a C$_{\rm 60}$ molecule  by pumping with light  carrying orbital angular momentum into the recently discovered super atomic molecular virtual orbitals. We predict a current generation leading to a magnetic pulse inside the cage in the range of mT. Such magnetic field are well detectable, e.g. by using
nitrogen-vacancy centers based magnetometry \cite{nv}.  The current can be tuned in sign and magnitude by changing the topological charge of the optical vortex. The pulse duration and intensity are additional tools for steering the generated magnetic field but also the underlying electronic structure can be exploited. Larger objects possessing highly degenerate SAMOs may accommodate much larger currents leading to higher magnetic fields.
{As the light frequency was tuned to SAMOs and hence far off the plasmon frequency, effects due to the induced fields play no relevant role here. The dynamics of plasmons driven by OAM  is currently in the scope of our research.} Summarizing, the  results point to a new non-invasive way of ultrafast optical manipulation of magnetically active endohedrals. In fact tailored fs magnetic pulses are shown to lead to ballistic switching of magnetization occurring  much faster than precessional switching \cite{sukhov2009local}.\\
%
%

This work was supported by the Deutsche Forschungsgemeinschaft under SPP 1840.

\end{document}